\documentclass[twocolumn]{aastex63}
\usepackage{graphicx, longtable}

\newcommand{\rearth}{$R_{\oplus}$}
\newcommand{\mearth}{$M_{\oplus}$}
\shorttitle{Radius and mass distribution of USP planets}
\shortauthors{Uzsoy et al.}

\begin{document}

\title{Radius and mass distribution of ultra-short period planets}

\author[0000-0001-9308-0449]{Ana Sof\'{i}a M. Uzsoy} 
\affiliation{Department of Physics, North Carolina State University, 421 Riddick Hall, Box 8202, Raleigh, NC 27695, USA}
\affiliation{Department of Astronomy \& Astrophysics, University of Chicago, 5640 S Ellis Ave, Chicago, IL 60637, USA}
\author[0000-0003-0638-3455]{Leslie A. Rogers}
\affiliation{Department of Astronomy \& Astrophysics, University of Chicago, 5640 S Ellis Ave, Chicago, IL 60637, USA}
\author[0000-0002-3286-3543]{Ellen M. Price}
\affiliation{Center for Astrophysics $|$ Harvard \& Smithsonian, 60 Garden Street, Cambridge, MA 02138, USA}

\begin{abstract}
Ultra-short period (USP) planets are an enigmatic subset of exoplanets defined by having orbital periods $<$ 1 day. It is still not understood how USP planets form, or to what degree they differ from planets with longer orbital periods. Most USP planets have radii $<$ 2 $R_{\oplus}$, while planets that orbit further from their star extend to Jupiter size ($>$ 10 $R_{\oplus}$). Several theories attempt to explain the formation and composition of USP planets: they could be remnant cores of larger gas giants that lost their atmospheres due to photo-evaporation or Roche lobe overflow, or they could have developed through mass accretion in the innermost part of the protoplanetary disk. The radius and mass distribution of USP planets could provide important clues to distinguish between potential formation mechanisms. In this study, we first verify and update the Kepler catalog of USP planet host star properties, incorporating new data collected by the Gaia mission where applicable. We then use the transit depths measured by Kepler to derive a radius distribution and present occurrence rates for USP planets. Using spherical and tidally distorted planet models, we then derive a mass distribution for USP planets.  Comparisons between the updated USP planet mass distribution and simulated planetary systems offer further insights into the formation and evolutionary processes shaping USP planet populations.

\end{abstract}


\section{Introduction}
\label{introduction}

Ultra-short period (USP) planets are an intriguing population of exoplanets with orbital periods $<$ 1 day. The first confirmed transiting USP planet was CoRoT-7b, with a radius of 1.68 $R_{\oplus}$ and an orbital period of 0.85 days \citep{Leger}. 

Since then, many additional USP planets have been discovered by the \textit{Kepler} spacecraft. During its 4-year mission, \textit{Kepler} monitored the brightness of around 200,000 stars.  In general, USP planets were not targeted by the \textit{Kepler} pipeline, and many searches of the \textit{Kepler} database utilized the Box-Least-Squares (BLS) method of detecting planetary transits. 
\cite{SanchisOjeda} (hereafter referred to as SO-14) applied Fourier transforms to the light curves of \textit{Kepler} target stars to reduce noise and isolate the periodic USP planetary transits. SO-14 presented 106 USP planet candidates whose radii fall almost exclusively below 2 $R_{\oplus}$, and found that USP planets occur around 1.10\% of M dwarfs and 0.83\% of K dwarfs. 
The \textit{K2} mission has nearly doubled the census of this enigmatic planet population \citep[e.g.][]{adamsjackson1,adamsjackson2, adams2020ultra}. \citet{adams2020ultra} identified 74 candidate USP planets in the first half of \textit{K2}. 

The USP planet population exhibits several characteristics that offer clues toward the planets' formation and evolution processes. The overall occurrence rate of USP planets is similar to that of hot Jupiters, though (in contrast to hot Jupiters) USP planets are more common around lower mass M dwarf stars than G and K dwarfs \citep{SanchisOjeda}. USP planets are typically small ($\lesssim 1.8~R_{\oplus}$). Among the USP planets with bulk density constraints, most are consistent with having an Earth-like compositions \citep{homogenous_analysis}, though a subset are iron-enhanced \citep{PriceRogers}. 
USP planets are typically accompanied by other planets with orbital periods between 1-50~days \citep{SanchisOjeda}, though the period ratio between USP planets and their nearest neighbors is generally $>4$ \citep{Steffen&Farr2013ApJ}, compared to the period ratios of 1.3 to 4 for adjacent planet pairs in the broader sample of Kepler multi-planet systems out to orbital periods of 100~days \citep{FabryckyEt2014ApJ}. USP planet have larger mutual inclinations than other planets among the \textit{Kepler} multi-planet systems  \citep{larger_mutual_inclinations}. The host stars of USP planets have a similar metallicity distribution to the host stars of hot sub-Neptunes (planet with radii $<4~R_{\oplus}$ and orbital periods between 1-10~days); unlike hot Jupiters, USP planets do not show an association with metal-rich stars \citep{Winnetal}. Finally, the ages of USP host stars are indistinguishable from the ages of field stars  \citep{HammerSchlaufman}, indicating that USP planets do not experience tidal inspiral while their host stars are on the main sequence.

In this study, we revisit the USP planet radius distribution. We update the radius distribution from SO-14 to include the improved stellar (and hence planetary) characterization and planet candidate vetting that has been completed in the intervening years. We focus on the USP planet sample discovered by \textit{Kepler} to leverage the injection-recovery completeness characterization performed by SO-14.

From the radius distribution, we derive the USP planet mass distribution, considering various assumed planet composition distributions. When computing model masses for the closest orbiting planets (with $P_{\rm orb}<10$~hours), we apply 3-dimensional interior structure models to take into account the tidal distortions of the planets' induced by their host stars. In Section~\ref{radii}, we report the updated radii of USP planets and their host stars. Section~\ref{occurrence} addresses the occurrence rate calculations, and in Section~\ref{massdistsection} we present the mass distribution. We discuss the implications of the inferred radius and mass distributions for the USP formation process in Section~\ref{discussion}, and summarize our findings in Section~\ref{conclusion}.

\section{Revised USP Planet \& Host Radii}
\label{radii}

This work aims to builds upon the USP planet characterizations of SO-14 by using updated stellar host parameters. The stellar population considered in this study matches that of SO-14, focusing on \textit{Kepler}-targeted G and K dwarfs, that is, stars with 4100 K $< T_{eff} <$ 6100 K and 4.0 $< log(g) <$ 4.9. We also limit the population to stars with Kepler magnitudes $m_{Kep} <$ 16 and with at least one quarter of data, leaving 97,940 stars. We identify 58 USP planet candidates in the SO-14 catalog with stellar hosts in the outlined population, eliminating ten candidates that have since been denoted false positives in the NASA Exoplanet Archive\footnote{\url{https://exoplanetarchive.ipac.caltech.edu/}}. 

We first revise the USP planet stellar host properties to incorporate updated stellar spectra and distance measurements collected since the original SO-14 catalog. We expand on the work of \citet{Winnetal} by including recent parallax ($\pi$) measurements from Gaia Data Release 2 (DR2), linked to their respective Kepler targets by Megan Bedell's \url{gaia-kepler.fun} crossmatch database.  Using the MIST isochrones \citep{isochrones}, we fit $K$ band magnitude, Gaia parallax ($\pi$) and, when available, spectroscopic parameters ($T_{eff}, log(g), [Fe/H]$) from the California Kepler Survey \citep{FultonPetigura, Winnetal} to stellar evolution models. Using MIST stellar isochrones \citep{isochrones}, we determine the host star's mass and radius.  SO-14's median host star radius uncertainties were +28\%, -8.3\%, while our median uncertainties are +2.5\%, -2.4\%.

\begin{figure}
  \includegraphics[width=\linewidth]{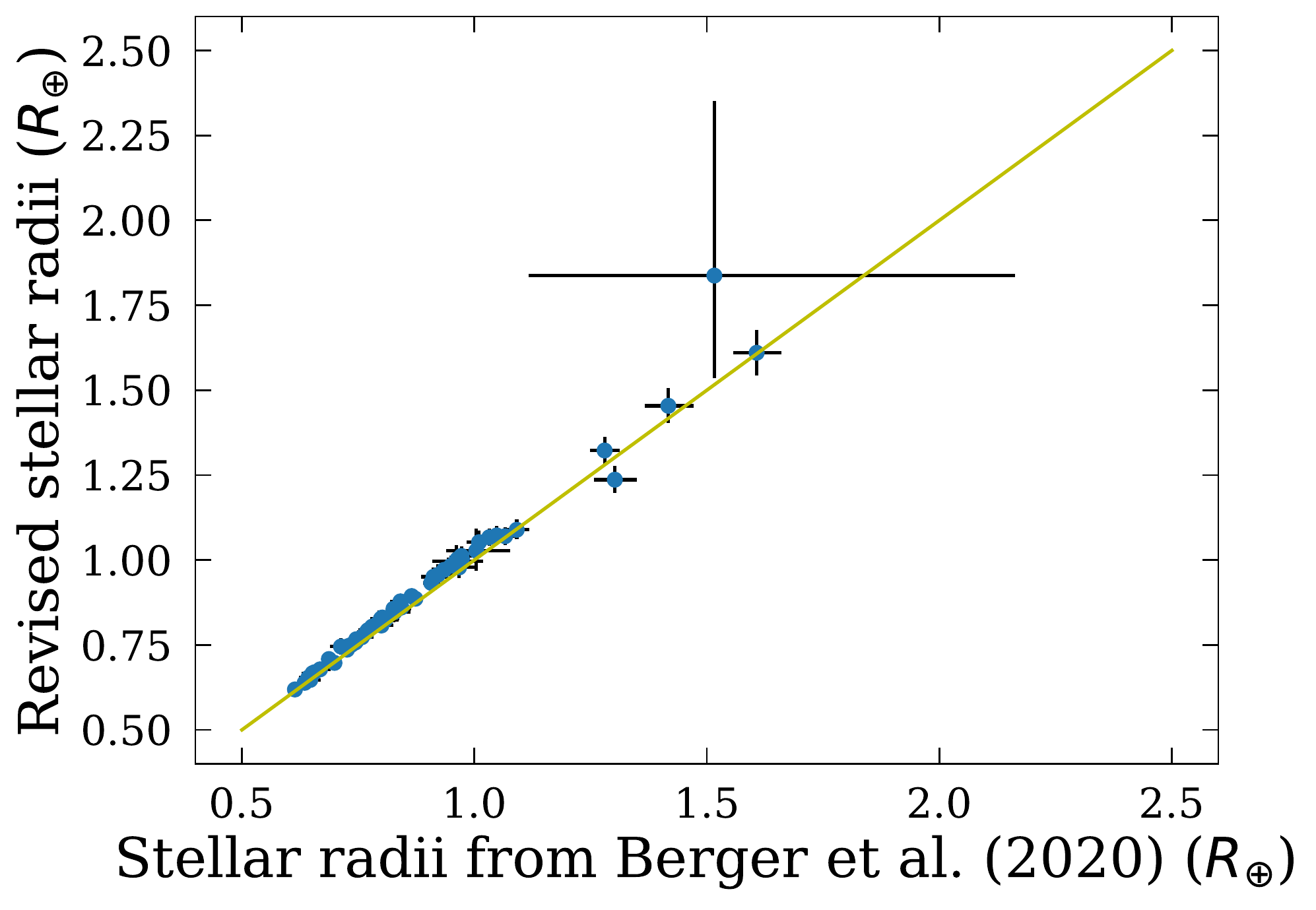}
  \caption{Comparison of USP planet stellar host radii from \citet{berger} and updated stellar radii calculated with {\tt isochrones} Python package using Gaia parallaxes and spectroscopic data from \citet{Winnetal}.  
  The yellow line represents $y = x$ for comparison.
}
  \label{fig:starcompareberger}
\end{figure}

Figure \ref{fig:starcompareberger} compares the stellar host radii calculated in this study to those used in \citet{berger}. The deviation between the values is likely created by the our use of host star spectroscopic parameters from \citet{Winnetal}. \citet{berger} uses Gaia parallaxes, stellar metallicities with an error of 0.15 dex, and stellar $g$ and $K_s$ photometry to derive $T_{eff}$, $log(g)$, radii, and masses for targets. We verify that these stellar radius calculations agree with those of \cite{berger} with a median deviance of 2.4\%.

We calculate the planetary radius, $R_{p}$, from the stellar radius, $R_{\star}$, using the same observed transit depths, $\delta$, as SO-14. For each planet, we take 1000 samples from both the stellar host radius posterior distribution calculated by \texttt{isochrones} and the transit depth posterior distribution (assumed to be normal), and calculate 1000 samples of planetary radii, using $R_{p} = \sqrt{\delta}R_{\star}$. 

\begin{figure}
  \includegraphics[width=\linewidth]{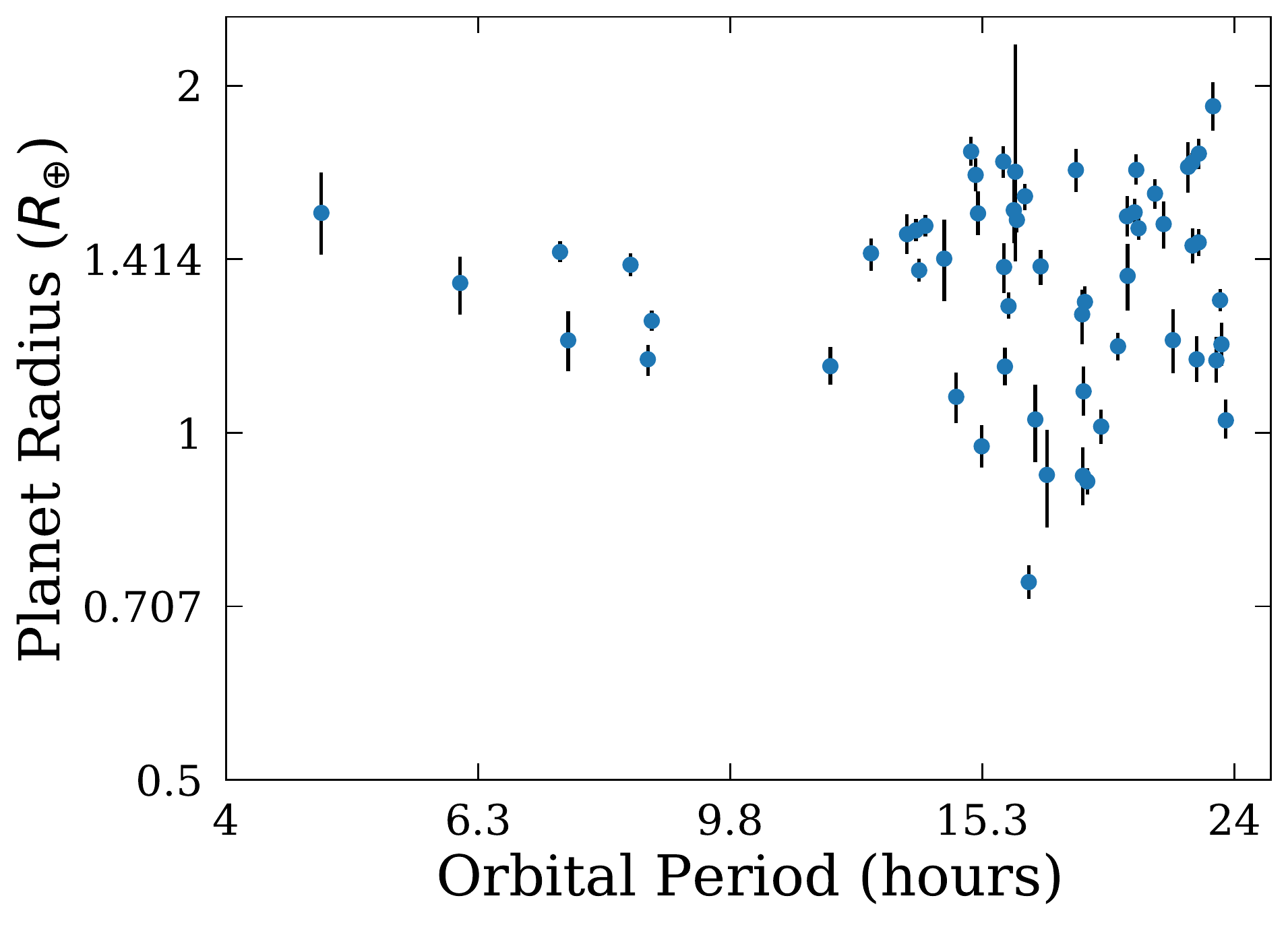}
  \caption{Median USP planet radius in logarithmic period-radius space. Error bars represent 1$\sigma$ (spanning from the 15.85th to 84.15th percentiles). The logarithmic scale matches that in SO-14. 
}
  \label{fig:radiidist}
\end{figure}

Figure \ref{fig:radiidist} shows the calculated radii of the 58 \textit{Kepler} USP planets. The majority of USP planets are ``super-Earth"-size, with their radii between 1.25 and 2 $R_{\oplus}$. Planets are more rare at shorter orbital periods, also noted in previous literature (SO-14; \citet{Howard2012}). The relative scarcity of USP ``Sub-Neptunes" (2 - 4 $R_{\oplus}$) and gas giants has also been noted previously by SO-14, suggesting that planets larger than 2 $R_{\oplus}$ cannot survive tidal forces and photoevaporation and remain intact as close to their host stars as smaller planets. 

\section{Occurrence Rate Calculations}
\label{occurrence}

Occurrence rates quantify how intrinsically common or rare planets are, accounting for both planets that are discovered and those that are missed. 

Factors such as observation bias, non-transiting geometries, and low signal-to-noise ratios lead to missed planets. In this section, we use the updated {\it Kepler} USP planet sample to calculate revised occurrence rates, accounting for undiscovered planets.

\begin{figure*}
  \includegraphics[width=\textwidth]{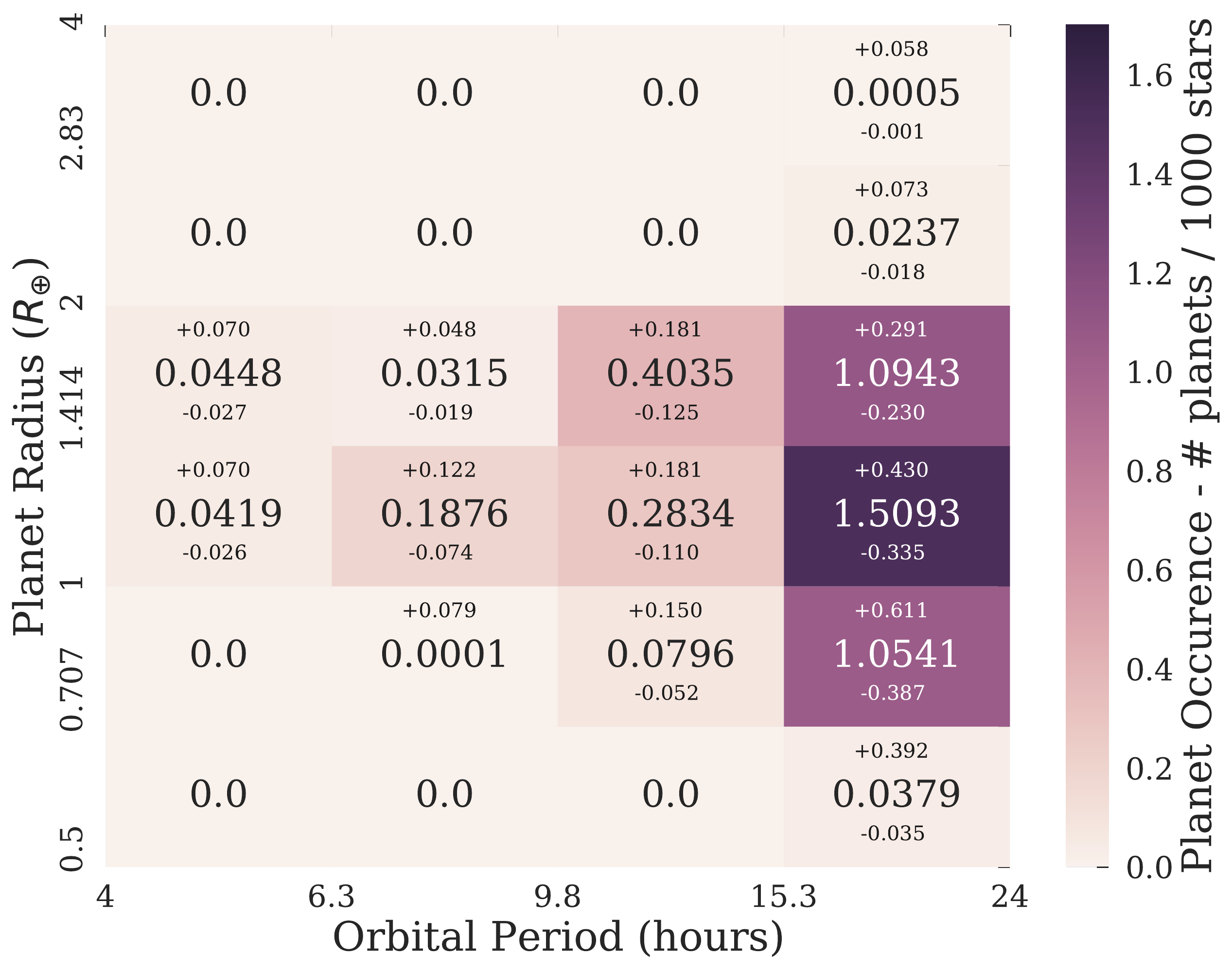}
  \caption{USP planet occurrence rates. Numbers within each cell represent occurrence rate (number of planets per thousand stars) as a function of radius and orbital period.  Cell color indicates the  occurrence rate of planets within the specified ranges of radius and orbital period on a logarithmic scale (using the same bins as \citet{Howard2012}). Uncertainties are shown above and below the occurrence rate for cells with nonzero occurrence rates.
}
  \label{fig:occurrencerates}
\end{figure*}

For transit surveys like Kepler, occurrence rates are often calculated within 2D bins in radius-orbital period space \citep[e.g.,][NASA Exoplanet Exploration Program Analysis Group (ExoPAG) Study Analysis Group (SAG) 13 Report\footnote{\url{https://exoplanets.nasa.gov/system/internal_resources/details/original/578_SAG13_standard_eta_definitions_v5.pdf}}]{Howard2012, PetiguraEt2013PNAS, Dressing&Charbonneau2015ApJ}. We follow an approach similar to that of \citet{Howard2012} to derive USP planet occurrence rates within 2D bins in radius-orbital period space. We deviate from the methods of \citet{Howard2012} in how we \textit{1)} account for measurement uncertainties on the planet radii \citep[following an approach similar to][]{SilburtEt2015ApJ, Dressing&Charbonneau2015ApJ}, \textit{2)} use the results injection-recovery tests from SO-14 to characterize the completeness of the USP planet detection pipeline instead of a SNR cutoff \citep[following an approach similar to, e.g.,][]{PetiguraEt2013PNAS}, and \textit{3)} compute uncertainties on the inferred occurrence rates. Ultimately, Poisson-likelihood-based hierarchical Bayesian approaches or Approximate Bayesian Computing  \citep[e.g.,][]{Youdin2011ApJ, ForemanMackey2014ApJ, NeilRogers, BrysonEt2021AJ} could be used to fit a parameterized planet distribution function and eliminate the need for binning when inferring occurrence rates, but these approaches are out of the scope of this work.

We follow an approach similar to that of \citet{Howard2012} to derive USP planet occurrence rates, considering 2D bins in radius-orbital period space. \cite{Howard2012} calculated the occurrence rate contribution of a planet with a given radius and orbital period as the ratio of the inverse transit probability and the number of stars from which that planet could have been detected.  The occurrence rate contribution $f_i$ for a planet $i$ of a specific radius and orbital period is given by:
\begin{equation}
    f_{i} = \frac{a_i/R_{*,i}}{N_{*,i}}
    \label{eq:fi}
\end{equation}
where $a$ is the planet's semi-major axis, $a/R_{*}$ is the inverse transit probability, and $N_{*,i}$ is the total number of stars in the survey around which the planet transit would have sufficient SNR to be detected. 

The total occurrence rate within a specified planet radius-orbital period bin, $f_{cell}$, is the sum of the individual occurrence rate contributions 
\begin{equation}
f_{cell} = \sum_{i = 1}^{n_{pl,cell}} \frac{a_i/R_{*,i}}{N_{*,i}}
\label{eq:fcell}
\end{equation}
of all $n_{pl,cell}$ planets detected within the 2D parameter space of the bin.

In order to calculate accurate occurrence rates, we must take into account the likelihood that a particular planet's transit signal would be sufficiently strong to be detected orbiting the stars in the survey. This is encapsulated by the factor $N_{*,i}$ in Equation~\ref{eq:fi}.

The signal-to-noise ratio (SNR) of a planet-star system measures the strength of the transit signal in comparison to the star's intrinsic noise. A higher SNR leads to a higher probability of a planet being detected. For each planet $i$, we calculated the SNR of the planet around each star $j$ in the survey, as follows:

\begin{equation}
    SNR_{ij} = \frac{\delta_{ij}}{\sigma_{CDPP,j}}\sqrt{\frac{T_{0,i}t_j}{(6\: hr)P_{orb,i}}}
\end{equation}
where $\delta$ is the transit depth, $\sigma_{CDPP}$ is the star's  CDPP (a measure of a its intrinsic noisiness) over six hours, $T_{0}$ is the transit duration, $t$ is the time spent actively observing the star (here the product of the data span and duty cycle), and $P_{orb}$ is the planet's orbital period in hours. 

To quantify the completeness of the USP planet search pipeline, SO-14 conducted transit signal injection-recovery tests in which they measured the probability of a planet being detected given its SNR. We fit a gamma CDF function to data from the lower right panel of Figure 7 in SO-14, which had $\chi^2$ = 0.026, as follows:

\begin{equation}
    P(x) = \frac{a}{\Gamma (b)} \gamma\left(b, \frac{x - c}{d}\right)
    \label{eq:gamma}
\end{equation}

\noindent where $P$ is the probability of detection, $x$ is the SNR, $a$ = 0.977, $b$ = 1.833, $c$ = 4.061, $d$ = 0.977, $\Gamma$ is the gamma function, and $\gamma$ is the lower incomplete gamma function.
Using Equation~\ref{eq:gamma}, we can quantify the probability of detection for each planet orbiting each star in the population of Kepler stellar targets with the characteristics described at the beginning of Section~\ref{radii}.

$N_{*,i}$ is the total number of \textit{Kepler} target stars around which a particular planet (indexed by $i$) can be detected. When calculating $N_{*,i}$, we use Equation~\ref{eq:gamma} to estimate the probability of a planet being detected around every star $j$ in our sample. We sum the probabilities as fractional stars to find the total number of stars around which that the planet could be detected:

\begin{equation}
    N_{*,i} = \sum_j P(SNR_{ij})
\end{equation}

After evaluating $N_{*,i}$ for each planet in our sample, we calculate $f_{cell}$ by summing the individual occurrence rate contributions of all the planets in the bin (Eqn.~\ref{eq:fcell}). We use 1000 radius samples for each planet, so each sample is weighted as $\frac{1}{1000}$ to take planet radius uncertainties into account.

The finite number of USP planets detected leads to inherent uncertainty in estimates of the occurrence rate, $f_{cell}$. 
In calculating uncertainties on $f_{cell}$, we consider the occurrence rate of \emph{transiting} planets within the radius-orbital period bin, calculated as:
\begin{equation}
   f_{tr,cell} = \sum_{i = 1}^{n_{pl,cell}} \frac{1}{N_{*,i}}.
\end{equation}
We model the detection of transiting planets within a specified radius-period bin as a binomial process with the occurrence rate, $f_{tr,cell}$, as the probability of ``success'' (i.e., the probability that a \emph{transiting} planet with specified properties is present around the star). We calculate uncertainties on $ f_{tr,cell}$ using the Wilson confidence interval to within 1$\sigma$ (15.85th to 84.15th percentiles):

\begin{equation}
    \frac{1}{1 + \frac{z^2}{n}}(\hat p + \frac{z^2}{2n}) \pm \frac{z}{1 + \frac{z^2}{n}}\sqrt{\frac{\hat p(1 - \hat p)}{n} + \frac{z^2}{4n^2}}
\end{equation}

\noindent where $z$ is the appropriate z-score (in this case, 1), $\hat p$ the observed proportion of successes (here $f_{tr,cell}$), and $n$ the effective number of trials (here $n_{*,eff,cell}=n_{pl,cell} / f_{tr,cell}$). The bounds of the confidence interval on $f_{tr,cell}$ are then scaled by $f_{cell}/f_{tr,cell}$ to set the uncertainties on the total occurrence rate of both transiting and non-transiting planets, $f_{cell}$. 

This method of calculating the uncertainties takes into account the fact that only transiting USP planets were detectable by \textit{Kepler}. The total number of planets successfully detected in each bin ($n_{pl,cell}$), thus depends most directly on the occurrence rate of transiting planets and not the total number of planets, transiting and non-transiting.

In this, we differ from the work of \citet{Howard2012} by calculating the effective number of stars searched as $n= n_{*,eff,cell} = n_{pl,cell} / f_{tr,cell}$ instead of $n_{pl,cell} / f_{cell}$, and the probability of success as $\hat p =f_{tr,cell}$ instead of $f_{cell}$.

Figure \ref{fig:occurrencerates} shows the planetary occurrence rates as a function of radius and orbital period. The general shape of Figure \ref{fig:radiidist} can be seen in the color gradient, with the majority of planets falling under 2 $R_{\oplus}$ with orbital periods close to the 1 day threshold. The most common USP planets have orbital periods between 15.3 and 24 hours and radii between 1 and 1.4 \rearth. They occur at 1.5 planets per thousand stars, a rate similar to that of hot Jupiters. When marginalized over planet radius, the values agree with those presented in SO-14 to within 2$\sigma$. 

We do not detect any statistically significant dependence of the USP planet radius distribution on orbital period within the \textit{Kepler} USP planet sample. 
To evaluate whether USP planet radii differ with orbital period, we perform two-sample KS tests comparing the radius distributions in each of the three lowest orbital period bins in Figure~\ref{occurrence} to that of the outermost bin. The resulting p-values of 0.36, 0.29, and 0.17 for the first, second, and third bins, respectively, show that the null hypothesis (that the planet radii within each orbital period bin were drawn from the same underlying planet radius distribution) cannot be ruled out.

\section{Planetary Mass Distribution}
\label{massdistsection}

\begin{figure}
  \includegraphics[width=\linewidth]{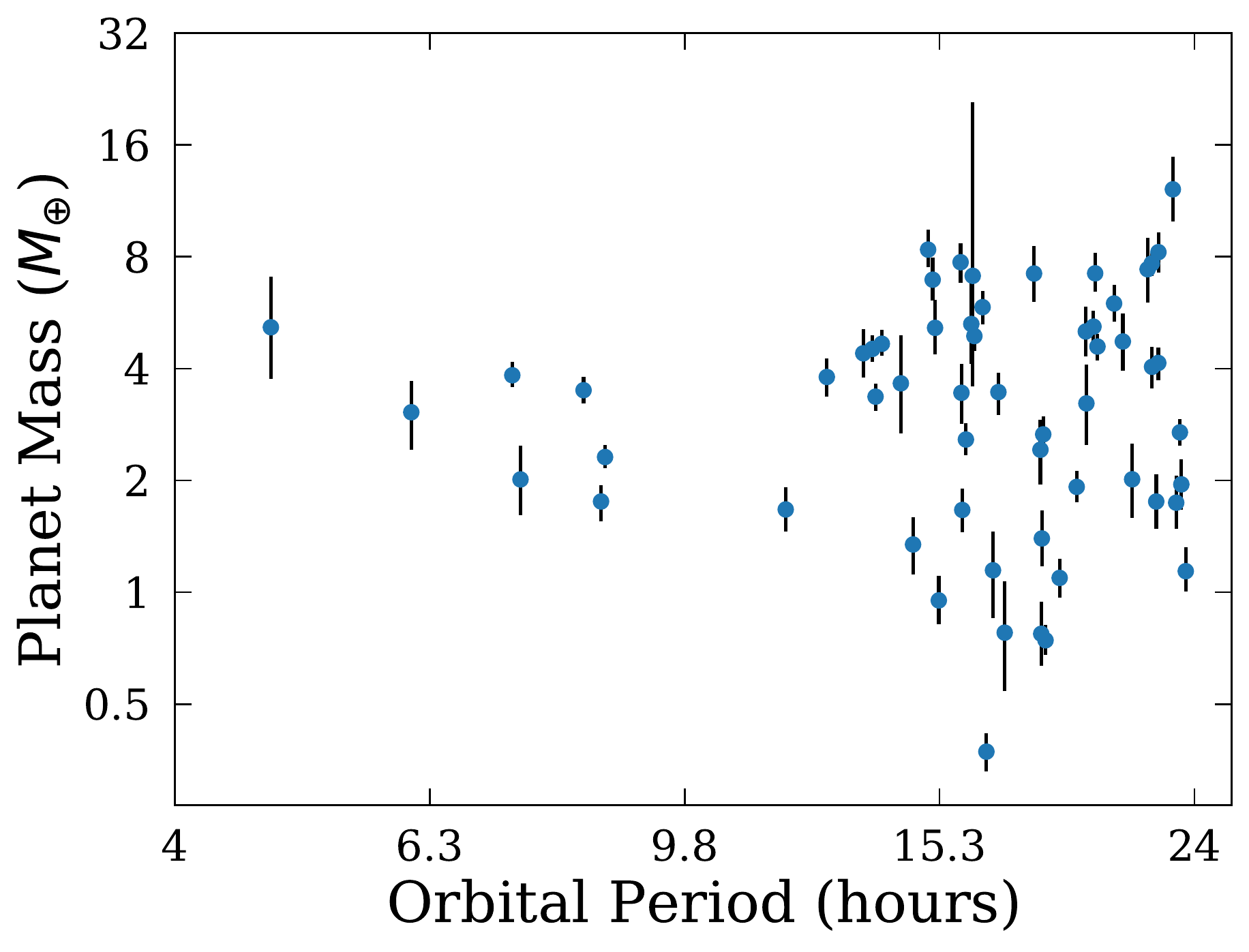}
  \caption{Median planetary masses, calculated using spherical integration assuming an Earth-like composition (CMF = 0.33). Error bars show uncertainties of 1$\sigma$ (15.85th to 84.15th percentile) 
}
  \label{fig:massdist}
\end{figure}

From the USP planet radius distribution, we derive a mass distribution using models of planet interior structure and various assumed composition distributions. 

The mapping from planet radius distribution to planet mass distribution depends on the composition distribution of USP planets. To address this, we calculate USP planet masses under a variety of composition assumptions: Earth-like, Mercury-like, and a random assortment of rocky compositions. Herein, we consider only rocky compositions (comprised of iron and silicates) for the USP planets. On USP orbits, primordial hydrogen-dominated envelopes would rapidly be lost to photoevaporation \citep[e.g.,][]{Owen&Wu2013ApJ, Lundkvist2016NatCo, Lopez}. Further, though planets on USP orbits could retain high-metallicity water-dominated envelopes, water-rich USP planets would commonly be $\gtrsim2~R_{\oplus}$, which is inconsistent with the observed USP planet radius distribution \citep{Lopez}. \citet{homogenous_analysis} found that most of the sample of 11 hot Earths with mass and radius measurements that they analyzed were consistent with an Earth-like composition, so we adopt the scenario wherein all USP planets have Earth-like compositions as our fiducial case. Since some USP are constrained to be iron-enhanced \citep{PriceRogers}, we also consider an extreme scenario wherein all USP planets have a Mercury-like composition to set an upper limit on the  planet masses, along with a scenario wherein each planet is randomly assigned an iron core mass fraction to quantify the effect of compositional diversity.

We use one-dimensional models of planet interior structure to convert from the USP planet radius distribution to the USP planet mass distribution in Section~\ref{sec:sphericalMR}. We include the effect of empirical mass measurements (for the subset of USP planets in our sample that have them) in Section~\ref{massmeasuredistsection}. Finally, we quantify the effect of tidal distortion on the USP planet mass distribution using three-dimensional models of planet interior structure in Section~\ref{sec:tidal}.

\subsection{USP Mass distribution from Spherical Models}
\label{sec:sphericalMR}

We first use spherical models to serve as a lower limit on the planet masses when orbital periods approach one day. The models are created by integrating the equation for the mass in a spherical shell, and the equation of hydrostatic equilibrium:
\begin{equation}
    \frac{dm}{dr} = 4\pi r^2 \rho
\end{equation}
\begin{equation}
        \frac{dp}{dr} = -\rho g,
\end{equation}
where $r$ is the distance to the planet's center, $m$ is mass interior to $r$,  $\rho$ is the density, and $p$ is pressure. We consider planets consisting of an iron core surrounded by a silicate mantle comprised of enstatite at pressures below 23~GPa, and perovskite phase MgSiO$_3$ at higher pressures. We use the same equations of state as \citet{PriceRogers} to determine the density at a specified pressure \citep{Seager+2007, ahrens1995mineral, karki, OLINGER1977325, Fe_epsilon}

We integrate in terms of $ln(p)$ starting from the planet center using the LSODA numerical integration scheme with a maximum step size of 0.1 and an absolute tolerance of $0.5\text{ x }10^{-6}$. Thus, given an input central pressure, $p_\mathrm{c}$, and core-mantle boundary pressure, $p_\mathrm{cmb}$, the total mass and radius of the planet is determined by the values of $m$ and $r$ at the surface, where $p=1$~Pa. 
The core mass and iron core mass fraction (CMF) are determined from the mass interior to the core-mantle boundary.

Using this integration procedure, we create planet interior structure models and compute planet mass, radius, and CMF over a grid of input core-mantle boundary pressures $p_\mathrm{cmb}$ and central to core-mantle boundary pressure ratios $\hat{p}_\mathrm{max}=p_\mathrm{max}/p_\mathrm{cmb}$. In our grid, we consider 64 equally spaced values of log($p_\mathrm{cmb}/1~\mathrm{Pa}$) ranging between 8 and 14  and 64 values of log($\hat{p}_\mathrm{max}$) 10 ranging from 0 to 0.1 and 54 ranging from 0.11 to 2. The additional resolution close to zero was necessary in order to interpolate masses for planets with very low CMFs. 

We interpolate within the grid of interior structure models to estimate the mass of each USP planet, for a specified CMF. For each of the 1000 radius samples of each planet, we perform a 2D linear interpolation to determine the values of $p_\mathrm{cmb}$ and $\hat{p}_\mathrm{max}$ that correspond to that particular combination of planet radius and CMF. We then use the pressure coordinates to interpolate a sample of the planet's mass for the specified iron core/silicate mantle composition. This distribution of 1000 mass samples can then be used to determine the median mass and uncertainty for each USP planet, listed in Table~\ref{masstable}.

Figure \ref{fig:massdist} shows the calculated masses for 58 USP planets, assuming they all have Earth-like bulk compositions (CMF = 0.33). Qualitatively, the distribution of modeled planet masses closely resembles the patter of planet radii shown in Figure \ref{fig:radiidist} as a function of orbital period, which is reasonable since the masses were calculated from the radii.

\begin{figure*}
  \includegraphics[width=\linewidth]{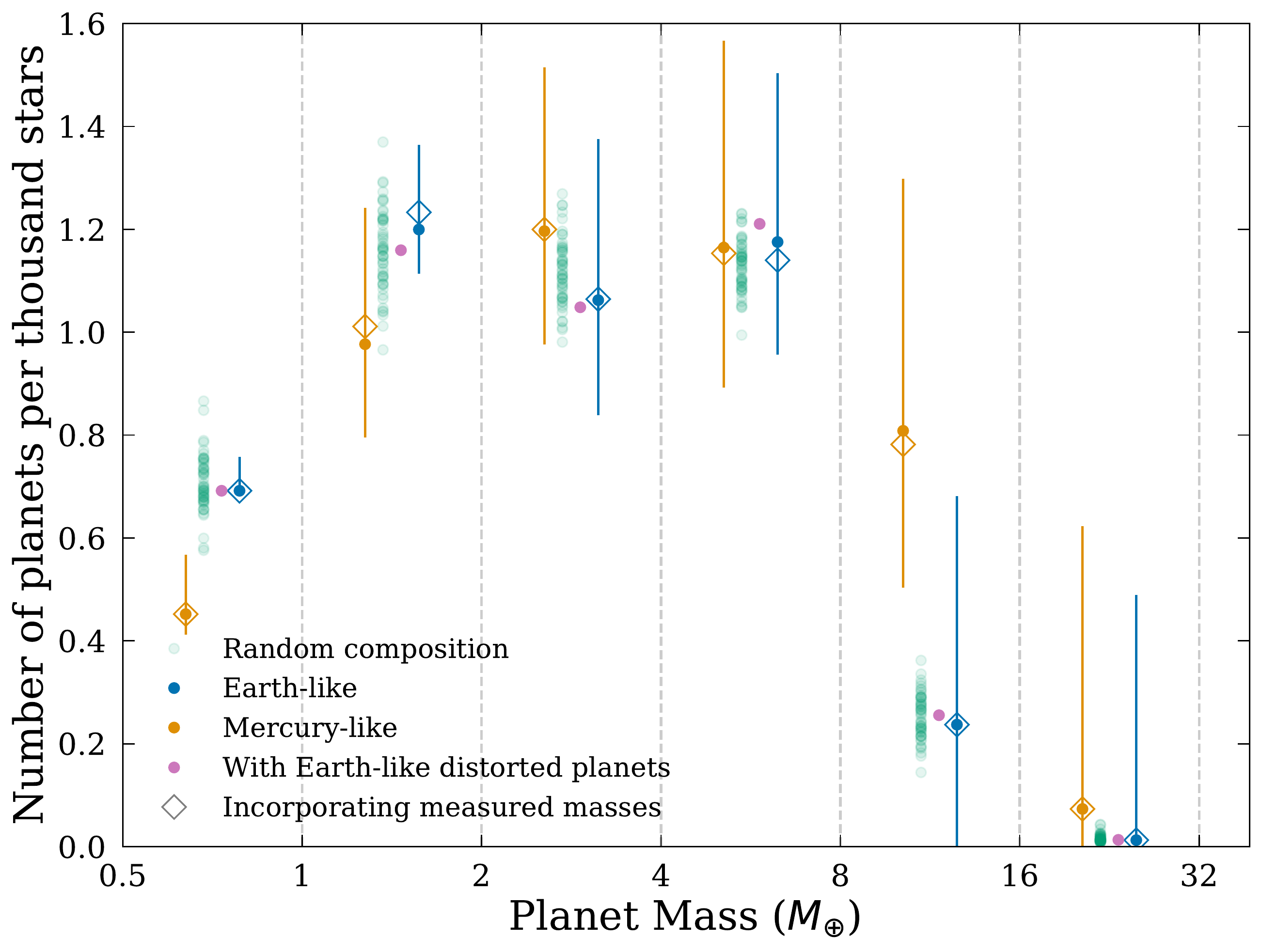} 
  \caption{Mass distributions with various assumptions of USP planet composition. Blue points are Earth-like with spherical planets (CMF = 0.33), pink points are Earth-like while including tidally distorted planets, orange points are Mercury-like with spherical planets (CMF = 0.7), green points are 50 mass distributions derived by randomly assigning to each planet a CMF drawn from Gaussian distribution with mean of 0.33 and standard deviation 0.1.  The open diamonds take into account measured masses as described in Section~\ref{massmeasuredistsection}. Error bars span the 15.85th to 84.15th percentiles on the occurrence rates for the Earth-like (blue) and Mercury-like (orange) mass distributions. Bins are logarithmically spaced with base 2. Points inside of each bin denote the inferred occurrence rate within the entire bin.
}
  \label{fig:massgraph}
\end{figure*}

We calculate the USP planet mass distribution in much the same way as the radius distribution in Section~\ref{radii}. For each planet, 1000 mass samples were used to calculate the occurrence rate in logarithmic mass-period bins, weigted by the occurrence rate contributions from Eqn~\ref{eq:fi}. Marginalizing over orbital period, we obtain a distribution of USP planet masses. 

Figure \ref{fig:massgraph} shows the distribution of spherical USP planet masses given several different composition assumptions - Earth-like (CMF = 0.33), Mercury-like (CMF = 0.7), and Random (normal distribution of CMFs with $\mu$ = 0.33 and $\sigma$ = 0.1). As is expected, the Mercury-like distribution yields more massive planets due to its higher iron content compared with the Earth-like distribution.  

To explore the effects of compositional diversity on the USP planet mass distribution, we randomly draw a core mass fraction for each planet from a Gaussian distribution with $\mu$ = 0.33 and $\sigma$ = 0.1 (truncating at zero). We repeat the random CMF assignments 50 times and plot the derived USP mass distributions for each (as green points in Figure \ref{fig:massgraph}). The plotting of multiple points illustrates the uncertainties in the USP mass distribution induced by the as-yet-unknown iron mass fraction distribution of the planets. The cumulative trend of the green points show a variety of possibilities for the true mass distribution of USP planets. 

Ultimately, the spread introduced by an intrinsic CMF scatter at the level of $\sigma$ = 0.1 is comparable to (at small masses) or smaller than (at large masses) the uncertainties in the USP mass distribution induced by the small number of USP planets in the {\it Kepler} data set. 

\subsection{Effect of Measured USP Planet Masses}
\label{massmeasuredistsection}

Two of the USP planets in our samples have measured masses (Kepler-78b and Kepler-10b) \citep{Kepler78b, Kepler10b}. To incorporate these measurements into our planet mass distribution, we replaced our mass samples for each of the planets in question (derived by transforming the planet radius samples with a model mass-radius relation for an assumed composition) with samples drawn from a normal distribution using the measured mass and its corresponding uncertainty as the mean and standard deviation. 

In Figure \ref{fig:massgraph}, the open diamonds show the Earth-like and Mercury-like distributions when measured masses are taken into account and the model masses are replaced with samples from the measured distribution as described above. The inclusion of measured masses shifts the mass distributions slightly towards lower planet masses, suggesting that Kepler-78b and Kepler-10b may have lower CMFs than our assumed values.

\subsection{Effect of Tidal Distortion}
\label{sec:tidal}

Given that USP planets orbit very close to their host stars, tidal forces can have significant impacts on the planets' morphology and thus the observed transit depths. Here, we turn to exploring the effect of tidal distortion on the inferred mass distribution of USP planet, using planet interior structure models that take the 3D tidally distorted shapes of the planets into account.

To model the 3D structures of USP planets, we use an updated version of the code presented in \citet{PriceRogers}, which uses the method of \citet{Hachisu1986a,Hachisu1986b} to self-consistently compute the shapes of ultra-short period exoplanets. As described in \citet{PriceRogers}, this iterative method takes as input the aspect ratio of the planet (measured at the origin of the coordinate system), the core-mantle boundary pressure $p_\mathrm{cmb}$, the ratio of central to core-mantle boundary pressure $\hat{p}_\mathrm{max}$, and the scaled distance from the planet to the star $\hat{a}$; it returns as output the parameters of interest such as planet mass, radii, and core mass fraction. Since the parameters of interest are output parameters, we cannot simply simulate a planet based on its mass and radii; instead, we interpolate within the model grid to find the parameters associated with a particular configuration.

\begin{figure}
  \includegraphics[width=\linewidth]{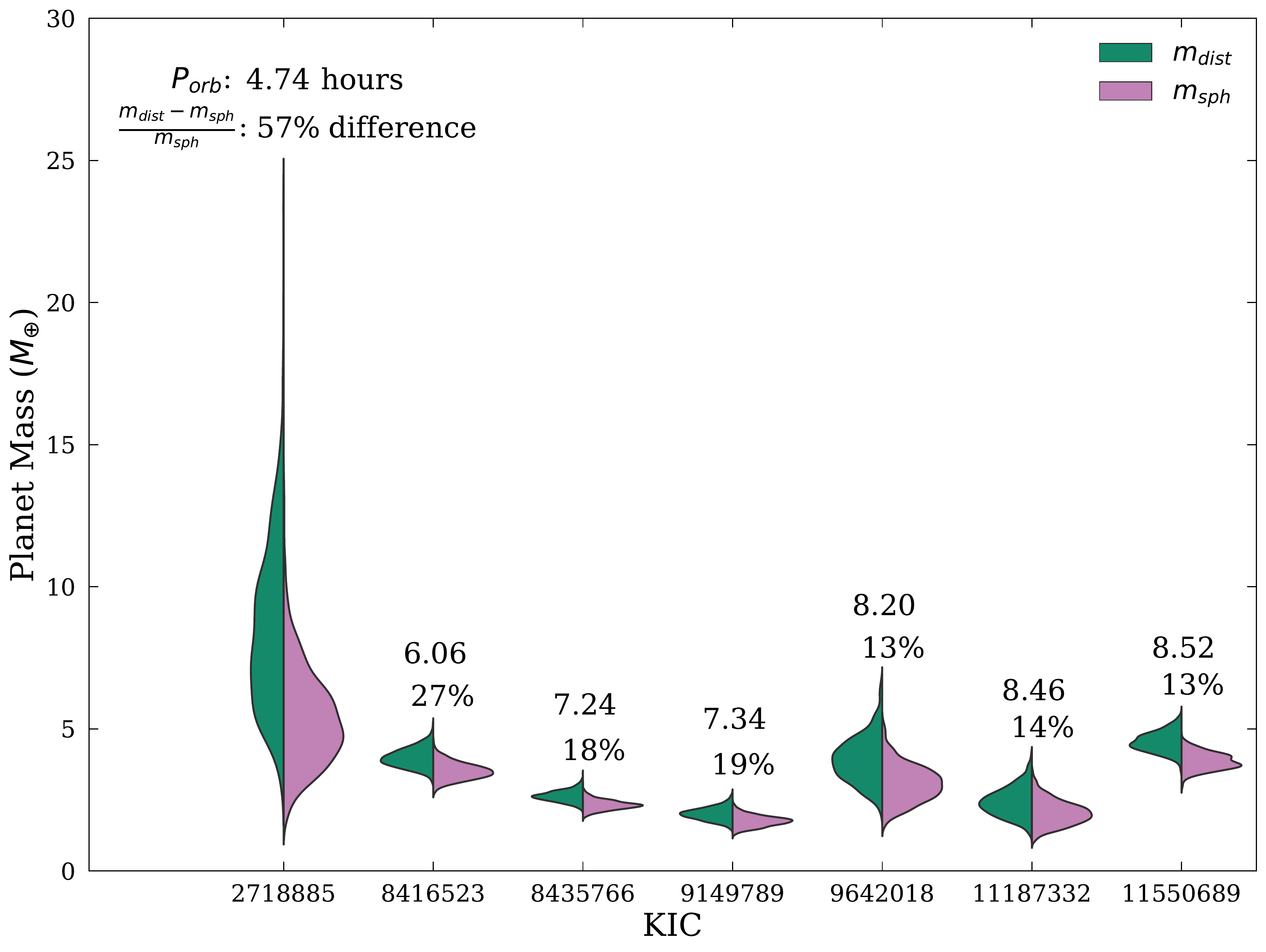}
  \caption{Comparison of planetary mass distributions for the seven USP planets with orbital periods less than 10 hours. The green distribution is for masses derived from tidally distorted models ($m_{dist}$), while pink is for masses derived from spherical models ($m_{sph}$). Each planet is annotated with its orbital period in hours and percent deviance between the median spherical and distorted masses, calculated as $\frac{m_{dist} - m_{sph}}{m_{sph}} * 100$. 
}
  \label{fig:distorted}
\end{figure}

For a specified transit depth, we expect tidally distorted masses to be larger than spherical masses, because the planet will be stretched towards its host start and thus have a larger volume than a sphere with the transit radius. Figure~\ref{fig:distorted} compares the masses inferred from the spherical and tidally distorted models for the seven USP planets with orbital periods below 10 hours. In each case, the median tidally distorted mass is larger than the median spherical mass, as was predicted. Additionally, the discrepancy between the distorted and spherical masses decreases as orbital period increases. This is also expected, as tidal forces are strongest for planets with shorter orbital periods.

The USP mass distribution incorporating the effect of tidal distortion (computed by replacing the masses derived from spherical models with the masses derived from 3D tidally distorted models) assuming all the planets have an Earth-like composition is shown by the pink points in Figure~\ref{fig:massdist}. The planet mass distribution  shifts towards higher masses relative to the  distribution based on spherical masses, while staying within the uncertainty on the spherical Earth-like distribution. The effect of tidal distortion on the inferred USP mass distribution is similar in magnitude to the effect of including the measured planet masses. 

In summary, tidal distortion can have a significant effect on the inferred masses of individual planets, increasing the modeled mass of the closest orbiting planets in the sample by up to 57\%. The most extreme USP planets ($P_{orb}\lesssim 10$~hours) that are significantly affected by tidal distortion are so rare, however, that the effect of tidal distortion on the inferred USP planet mass distribution is diluted.

\section{USP Planet Formation Implications}
\label{discussion}

\begin{figure*}
  \includegraphics[width=\linewidth]{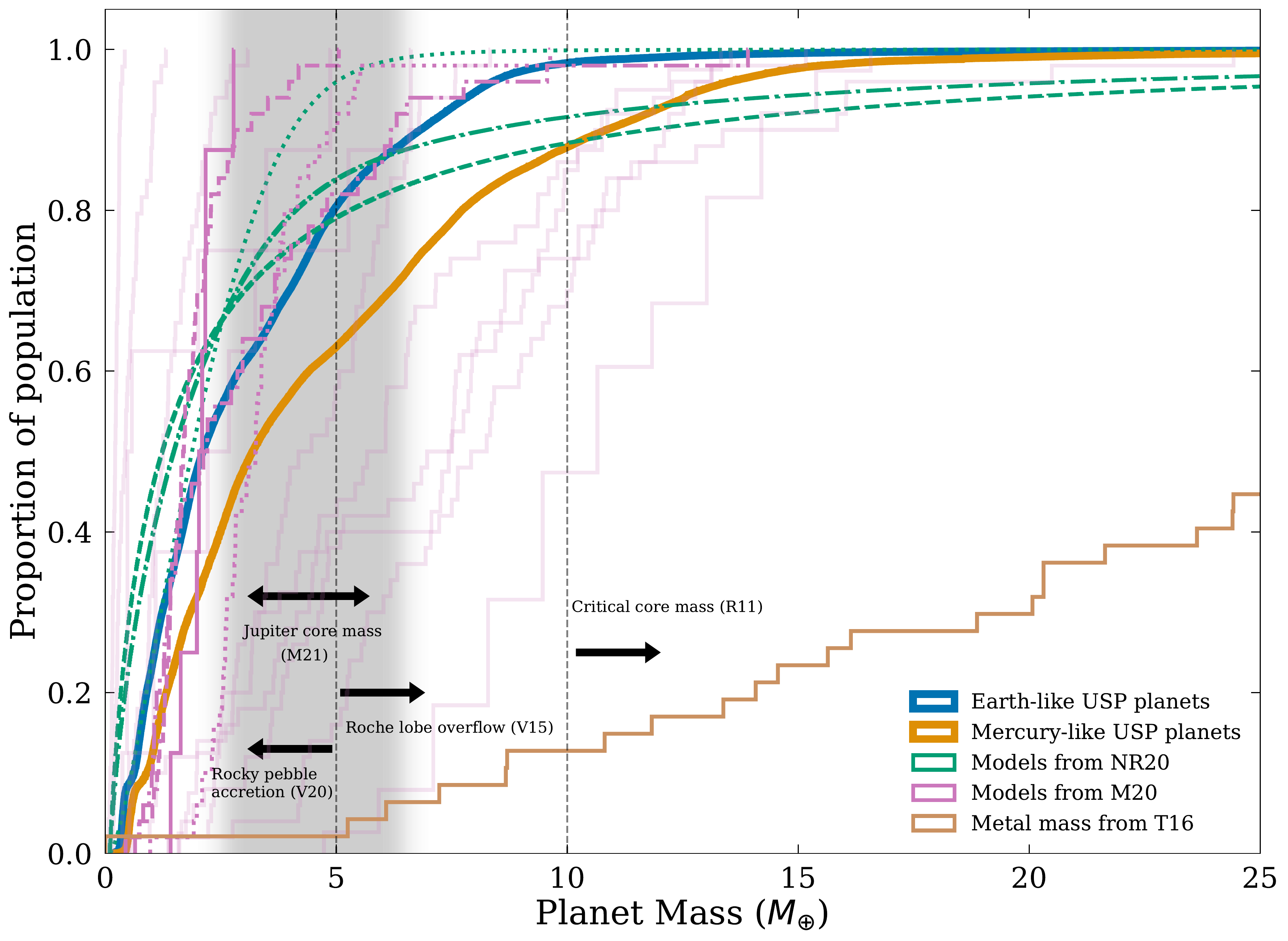}
  \caption{Cumulative distribution functions of various potential USP mass distributions. The thick blue ``Earth-like" and orange ``Mercury-like" lines correspond to the distributions described in Section~\ref{massdistsection}, weighted by occurrence rate contribution. The green lines correspond to the USP planet distributions from \citet{NeilRogers} (abbreviated as NR20): intrinsically rocky (dashed), evaporated cores (dotted) and the total combined population of rocky planets (including both intrinsically rocky and evaporated cores, dash-dot). The pink transluscent lines show innermost planet masses from all populations in \citet{Genesis} (abbreviated as M20), with the pink solid, dashed, dotted, and dash-dotted lines corresponding to the innermost planet masses from the ``Gen-P", ``Gen-O-s22", ``Gen-HM", and  ``Gen-M-s22" populations from M20, respectively. The brown line corresponds to the distribution of total heavy-element ``metal" mass, $M_Z$, inferred for the gas giant planets from \citet{Thorngren} (abbreviated as T16). 
  The arrows correspond to constraints from planet formation models; super-Earths formed by rocky pebble accretion have masses under 5 \mearth~(\citet{5Mearth}, abbreviated as V20), USP planets under 5 \mearth cannot be formed by mass loss through dynamically stable Roche lobe overflow (\citet{Valsecchi_2015}, abbreviated as V15), and cores of gas giants formed by runaway gas accretion had traditionally been thought to have a critical (minimum) mass of 10 \mearth \citep[][R11]{Rafikov}. The gray shading indicates Jupiter's central condensed core mass \citep{MiguelEt2021submitted}.}
  \label{fig:compareformation}
\end{figure*}

Our updated USP planet radius distribution, and inferred USP planet mass distribution provide useful empirical anchors to which the predicted outcomes of USP period planet formation mechanisms can be compared. Figure ~\ref{fig:compareformation} presents the cumulative mass distributions functions (CMDF) inferred for the \textit{Kepler} USP planets, along with constraints from various planet formation scenarios.

There are various formation mechanisms and evolutionary processes that may contribute to the observed population of USP planets. These mechanism include 
inward migration of rocky and sub-Neptune planets \citep[e.g.,][]{Mardling&Lin2004ApJ, LeeandChiang, PetrovichEt2019AJ, pulai, millholl2020formation}, photo-evaporation of Neptune and sub-Neptune planets \citep[e.g.,][]{Owen&Wu2013ApJ, Lundkvist2016NatCo, Lopez}, and Roche-lobe overflow (RLO) of gas giant planets \citep[e.g.][]{Valsecchi_Rasio_Steffen, Valsecchi_2015,Jackson, Konigl}.

The defining characteristic of USP planets is their short orbital period. Because orbital periods of $\leq 1$~day fall interior to the dust-sublimation radius in their nascent protoplanetary disks, USP planets must have experienced inward migration either during formation or later dynamical evolution. USP planet formation scenarios involving low-mass (rocky or sub-Neptune) progenitors entail inward migration driven by tidal dissipation. Possible sources of tidal dissipation include tides raised on the host star \citep{LeeandChiang}, planet eccentricity tides driven by secular planet-planet interactions \citep{PetrovichEt2019AJ,pulai}, and planetary obliquity tides driven by Cassini states with forced non-zero obliquities \citep{millholl2020formation}.

In each of these formation pathways, the mass distribution of observed USP planets that we have derived would reflect the mass distribution of the planets (and/or the cores of the planets) at the initial orbital separations from which USP planets are sourced \citep{millholl2020formation}: near the inner-edge of the magnetospherically truncated protoplanetary disk in the case of \citet{LeeandChiang}, orbital periods of $\sim5-10$~days in the high-eccentricity migration scenario of \citet{PetrovichEt2019AJ}, orbital periods of $\sim1-3$~days in the moderate-eccentricity migration scenario of \citet{pulai}, and orbital periods of $\sim1-5$~days in the obliquity-driven tidal runaway scenario of \citet{millholl2020formation}.  

\subsection{Low-Mass USP Planet Progenitors}

After migrating inward toward their host star, any initially Neptune and sub-Neptune-sized USP planet progenitors with primordial envelopes, would be subject to photo-evaporation by the intense levels of photoionizing radiation at their new orbital separations \citep[e.g.,][]{Owen&Wu2013ApJ, Lundkvist2016NatCo, Lopez}. The sub-Neptunes would rapidly lose their envelopes, leaving remnant evaporated USP cores. The H/He envelope of a sub-Neptune planet represents a significant fraction of the planets radius, but only a small fraction of the planet's mass \citep[typically $\sim1\%$ by mass][not accounting to hydrogen dissolved in the core]{Wolfgang&Lopez2015ApJ}. Thus, though the radius distribution of USPs may not reflect the radius distribution of their progenitors (due to photo-evaporation), the USP mass distribution will reflect the mass distribution of both the progenitor's cores, and the sub-Neptune-mass progenitors themselves.

\citet{NeilRogers} used hierarchical Bayesian modeling along with parameterized multi-component mixture models for close-orbiting planet populations (accounting for photo-evaporation) to constrain the joint mass-radius-period distribution of \textit{Kepler} transiting exoplanets. The mass distributions of USP planets inferred by \citet{NeilRogers} using their preferred model (their ``Model 3") are displayed in Figure~\ref{fig:compareformation}. ``Model 3" from \citet{NeilRogers} comprises a mixture of 3 planet sub-populations: gaseous planets with primordial envelopes, evaporated cores (the remnant rocky cores of formerly gaseous planets that lost their envelopes to photo-evaporation due to their short envelope-loss timescales), and intrinsically rocky planets (planets that initially formed as rocky planets are whose mass and period distributions are not tied to the gaseous planet population). The planet mass distributions from \citet{NeilRogers} were generated by sampling from the posterior predictive distribution obtained from their fit of ``Model 3" to the Kepler DR25 catalog (including radial velocity planet mass measurements, when available). In Model 3, USP planets with radii below 2 $R_{\oplus}$ either intrinsically rocky or evaporated cores. We present both the total USP planet CMDF from \citet{NeilRogers}, along with the CMDFs for the evaporated core and intrinsically rocky sub-populations. Overall, the evaporated cores from \citet{NeilRogers} tend to have masses below 5~\mearth, while the mass distribution of intrinsically rocky planets is broader and extends up to higher masses.

Our Earth-composition USP mass distribution is not significantly different from the combined (evaporated core and intrinsically rocky) mass distribution from \citet{NeilRogers}.  We use the 2-sample KS test to compare our fiducial (Earth-like composition) \textit{Kepler} USP planet mass distribution with the combined distribution from \citet{NeilRogers} and find that the null case that both samples were drawn from the same underlying distribution cannot be rejected with the KS-test ($p$ = 0.108). In contrast, the mass distributions of the two individual sub-populations from \citet{NeilRogers} are significantly different from our Earth-composition USP planet mass distribution ($p < 0.05$); we obtain $p=0.013$ and $p=0.014$ when comparing the USP planet mass distribution to the evaporated cores and intrinsically rocky distributions, respectively.

Comparing our inferred USP mass distribution to the outcomes of planet formation simulations, offers some insights into the plausibility of various planet formation pathways for USP planet progenitors.

\citet{5Mearth} simulated planetary formation by pebble accretion inside the ice line, and found that most cores formed by dry pebble accretion are less than 5 $M_{\oplus}$. 
In our sample, 41 of the 58 planets have median masses below 5 $M_{\oplus}$ (under the assumption of an Earth-like composition), suggesting that dry pebble accretion is a viable initial formation mechanism for the rocky planets now found at USPs \citep{5Mearth}.

\citet{Genesis} used n-body simulations to model the final stages of planet formation (when planets accrete from planetary embryos) for a variety of initial conditions and used the Exoplanet Population Observation Simulator (EPOS) to compare them to \textit{Kepler} systems. \citet{Genesis} performed 15 sets of formation simulations, wherein each set consisted of multiple (typically 50) runs for the same choice of key input parameters (describing the disk solid-surface density profile, disk inner and outer radii, and the presence/absence of nebular gas and planetesimals). We compute a CMDF for each set of simulations, including only the mass of the innermost (shortest orbital period) planet from each run, since the innermost planet of each system is the most likely to evolve into an USP planet. We  display in Figure~\ref{fig:compareformation} the CMDFs obtained from all 15 sets of formation simulations.  The 15 distributions span a wide range of parameter space both above and below our USP planet mass distributions, but all have masses lower than 20 \mearth.

We highlight in Figure~\ref{fig:compareformation} run sets for four different accretion scenarios that produce planets in the size range from 1-3\rearth  and orbital period range from 5-100 days \citep[the same runs sets highlighted in Figures 9 and 10 of][]{Genesis}. Run set ``Gen-HM" consists of an embryo-only scenario; run set ``Gen-M-s22" includes nebular gas and takes orbital migration/damping into account; run set ``Gen-O-s22" includes an extended disk model; and run set ``Gen-P" includes planetesimals. 
We performed 2-sample KS tests comparing the ``Gen-HM",``Gen-M-s22", and ``Gen-O-s22" CMDFs to our Earth-like USP planet mass distribution. ``Gen-P" was not included in this analysis since it comprises only 8 runs (and thus its CMDF is based on only 8 simulated planets). Our Earth-like USP distribution was significantly different from the ``Gen-HM" distribution ($p$ = $2.67 \times 10^{-5}$) and the``Gen-O-s22" distribution ($p$= 0.006). In contrast, the ``Gen-M-s22" distribution was not significantly different from the Earth-like USP planet distribution ($p$ = 0.276) suggesting that orbital migration could play a key role in USP planet formation.

\subsection{Gas Giant USP Planet Progenitors}

One scenario for the origin of USP planets is that they are the exposed cores of hot Jupiters formed by core accretion. The similarity between the occurrence rates of USP planets and hot Jupiters \citep[as noted by ][]{SanchisOjeda,Steffen&Coughlin2016PNAS} could be interpreted as 
indirect evidence for 
a connection between these two populations. However, the significant differences between the metallicity distributions of hot Jupiter host stars \citep[which are preferentially metal-rich][]{Fischer&Valenti2005ApJ} and USP planet host stars \citep[which show no preference to be metal-rich][]{Winnetal},  constrains the fraction of USP planets that can be remnant hot Jupiter cores \citep[no more than 36\% at 2$\sigma$ confidence][]{Winnetal}. 
Even though no more than about a third of USP planets can be the vestiges of former Jupiters, the possibility of using a sub-population of USP planets to glimpse the  heavy element core distribution of hot Jupiters is intriguing.

How could a hot Jupiter lose its H-He envelope and leave a remnant core? Photo-evaporation alone is insufficient to remove the massive (order-of-magnitude $10^2~M_{\oplus}$) envelope of a gas giant planet \citep{Murray-ClayEt2009ApJ}. Hot Jupiters could, however, lose their envelopes through Roche-lobe overflow (RLO), if the planet's orbit crosses the Roche limit \citep[e.g.,][]{Valsecchi_Rasio_Steffen, Valsecchi_2015, Jackson, Ginzburg&Sari2017MNRAS}. \citet{HammerSchlaufmann2019AJ} has found strong evidence that hot Jupiter planet orbits tidally decay toward their host star while the star is on the main sequence; by comparing galactic velocity dispersions, \citet{HammerSchlaufmann2019AJ} showed that hot Jupiters host stars are on average younger than similar field stars without hot Jupiters.

\citet{Valsecchi_2015} modeled the coupled orbital and mass-loss evolution of hot-Jupiters that reach their Roche limits and undergo dynamically stable RLO mass transfer, finding an anti-correlation between the mass and the orbital period of the remnant cores left-over. \citet{Valsecchi_2015} inferred that stable RLO from hot Jupiters cannot account for USP planets with masses less than $\lesssim5$~\mearth, which rules out this evolution pathway for roughly 70\% of the \textit{Kepler} USP planets (assuming an Earth-like composition).

If the mass loss from hot Jupiters via RLO is unstable on dynamical timescales \citep[instead of stable as][assumed]{Valsecchi_2015}, remnant cores of tidally stripped hot Jupiters may still plausibly comprise a significant  sub-population of USP planets \citep{Konigl}. 
The mass loss from hot Jupiters via RLO will be dynamically unstable if the escaping mass from the planet falls directly onto the star (where its angular momentum contributes to spinning up the star), instead of forming an accretion disk that can torque the orbit of the planet \citep{Jia&Spruit2017MNRAS}. \citet{Konigl} performed Monte Carlo experiments to demonstrate that the remnant cores left by unstable RLO from Jupiters brought into the vicinity of their Roche limits by high-eccentricity migration can plausibly account for the planet radius and orbital period distribution of the dynamically-isolated ``hot Earths" with $P_{\rm orb}\sim 1$~day identified by \citet{Steffen&Coughlin2016PNAS}. In \citet{Konigl}'s simulations, most of the remnant cores with USPs ($P_{\rm orb}<1$~day) correspond to ``late cores" of Jupiters that cross the Roche limit not on their original high-eccentricity orbit but only after experiencing orbital circularization and tidal decay.

 Figure~\ref{fig:compareformation} compares the mass distributions inferred for USP planets to various predictions and measurements for the core masses of gas giant planets. \citet{Thorngren} used planet interior structure models to infer the heavy-element masses of 47 giant planets with mass and radius measurements. In their models, \citet{Thorngren} placed up to $10~M_{\oplus}$ in a central heavy element core, incorporating any remaining heavy element (or ``metal") mass in the H-He-dominated envelope. Typical USP masses are significantly lower than both the total heavy element masses and the modeled core masses of the gas giants in the \citet{Thorngren} sample. A 2-sample KS-test confirms the stark differences between the \citet{Thorngren} and USP planet CMDFs ($p<10^{-18}$). The critical (i.e., minimum) core mass for runaway gas accretion is canonically inferred to be $\gtrsim 10~M_{\oplus}$ \citep[][]{Rafikov}, which is more massive than 97\% of our inferred USP planet mass distribution (assuming an Earth-like composition).

If a significant faction ($\gtrsim10\%$) of USPs are the remnant cores of tidally stripped hot Jupiters, the inferred USP mass distribution implies that gas giant cores are typically smaller than $7.6~M_{\oplus}$, the 90th percentile of the Earth-like USP CMDF. 
This result aligns with recent revelations from NASA's \textit{Juno} mission \citep[launched in 2011,][]{juno}, which is orbiting Jupiter, studying its inner structure. The core of Jupiter is revealed to be ``fuzzy'', likely indicating a composition gradient from the solid center to the gaseous H/He envelope \citep{WahlEt2017GeoRL, MiguelEt2018A&A}. A recent analysis of \textit{Juno} data by \citet{MiguelEt2021submitted} constrains the mass of Jupiter's inner compact core to be approximately between 2 and 7 \mearth.

The critical core mass may be lower than the textbook value of 10~\mearth. 
Dust depletion in the planet forming disk may decrease the critical core mass for runaway accretion \cite[][]{Rafikov, PisoEt2015ApJ, HoriEt2010ApJ} by lowering the opacities and facilitating the cooling and contraction of the proto-envelope.

As \textit{Juno} has confirmed, gas giants likely do not have well-defined divisions between a heavy-element core and H/He envelope, instead having ``fuzzy" cores and a gradual increase in the heavy-element mixing ratios toward the planet center \citep{Helled&Stevenson2017ApJ}. Multiple processes may contribute to gas giant core ``fuzziness". 
During a gas giant's formation, dissolution of accreting rocky planetesimals in the gaseous envelope of the planet, can lead to substantial metal enrichment of the envelope (at the expense of the central compact core), resulting composition gradients within the planet \citep{BodenheimerEt2018ApJ,Valletta&Helled2020ApJ}. An energetic head-on giant impact between a proto-gas giant and a $\sim 10~M_{\oplus}$ planetary embryo can also disrupt the gas giant's primordial compact core and mix the heavy elements within the high-pressure envelope, leading to a dilluted core \citep{LiuEt2019Nature}. 
Even after a gas giant's initial formation, the initial heavy element core of the planet may erode over time as the planetary materials -- including H$_2$O \citep{Wilson&Militzer2012ApJ}, Fe \citep{WahlEt2013ApJ}, MgO \citep{Wilson&Militzer2012PRL}, and SiO$_2$ \citep{Gonzalez-CataldoEt2014ApJ}  --- are soluable in metallic hydrogen at the high-pressure conditions at the center of a Jupiter-mass planet. Further study is needed to assess how much mass would be left over in an USP remnant if a gas-giant planet with a ``fuzzy" core loses its gaseous envelope to RLO.  

\section{Summary \& Conclusion}
\label{conclusion}

We present an updated radius distribution and a completeness-corrected mass distribution of the \textit{Kepler} USP planets. Expanding on the work of \citet{SanchisOjeda}, we take into account new parallax \citep{Gaia2018} and spectroscopic \citep{Winnetal} measurements to calculate the stellar host and USP planet radii, and present the occurrence rates of USP planets over logarithmic radius-period bins. 
We then calculate USP mass distributions using various assumptions for the planet compositions, taking into account the tidal distortion of planets with orbital periods less than 10 hours and USP planet mass measurements (when available). 

We explore the implications of our results for USP planet formation mechanisms and evolutionary processes. 
Comparisons to models from \citet{NeilRogers}, \citet{Genesis}, and \citet{5Mearth} suggest that USP planets could be formed by a variety of mechanisms, including photoevaporation of sub-Neptunes, accretion in the presence of orbital damping from gas, and dry pebble accretion. In contrast, formation scenarios that predict planet masses that are significantly higher or lower than our inferred USP planet mass distribution are disfavored. If  a  significant  sub-population  ($\gtrsim10\%$)  of  USP's are  the  remnant  cores  of  tidally  stripped  hot Jupiters \citep[as suggested by][]{Konigl}, the inferred USP mass distribution implies  that  gas  giant  cores  are  typically  smaller than 7.6\mearth (the 90th percentile of the Earth-like USP planet CMDF).
Typical  USP planet masses are significantly lower than the total heavy element masses of gas  giants exoplanets \citep{Thorngren}, but are comparable to recent constraints on the mass of Jupiter's central compact core \citep[][]{juno, MiguelEt2021submitted} accounting for a gradient of heavy-elements dissolved in the envelope.

In this work, we have focused on the USP planet sample discovered by \textit{Kepler}, leveraging the injection-recovery completeness characterization performed by SO-14. With only 58 planet candidates in the sample, however, small number statistics are a dominant source of uncertainty in our derived USP planet mass distribution (swamping the effect of planet RV measurements, tidal distortions, and CMF variance of the order of 0.1). \textit{K2} and \textit{TESS} photometry is enabling new USP planet discoveries around brighter stars (on average) than that SO-14 \textit{Kepler} sample that are more amenable to follow-up characterization: e.g., K2-141b \citep{BarraganEt2018A&A, MalavoltaEt2018AJ}, HD~213885~b \citep{EspinozaEt2020MNRAS}, HD~80653~b \citep{FrustagliEt2020A&A}, and TOI-561 b \citep{WeissEt2021AJ, LacedelliEt2021MNRAS}. 
Once the completeness and reliability of of the \textit{K2} and \textit{TESS} transit surveys have been fully characterized, the USP mass distribution derived here could be further improved by aggregating USP planet samples and occurrence rates from \textit{Kepler}, \textit{TESS}, and \textit{K2}. Additional USP planet mass measurements will further help to benchmark our assumed CMFs and M-R relations used to map from the radius distribution to the mass distribution.

Overall, continued refinement of the radius and mass distribution of USP planets may enable further disambiguation of the formation and evolutionary processes sculpting this intriguing class of exoplanets.

\acknowledgments
This work made use of the gaia-kepler.fun crossmatch database created by Megan Bedell.

Some of the data presented in this paper were obtained from the Mikulski Archive for Space Telescopes (MAST). STScI is operated by the Association of Universities for Research in Astronomy, Inc., under NASA contract NAS5-26555.

The authors would like to thank BJ Fulton for providing data from \citet{FultonPetigura}, Yamila Miguel for providing data from \citet{MiguelEt2021submitted}, Travis Berger for providing details regarding \citet{berger}, and Andrew Neil for helpful discussion and providing data from \citet{NeilRogers}. 

 L.A.R. acknowledges NSF grant AST-1615315. This work was completed in part with resources provided by the University of Chicago's Research Computing Center. A.S.M.U acknowledges support by a NC State Park Scholarship and a Goldwater Scholarship.

\bibliography{exoplanets}

\begin{deluxetable*}{cccccc}
\tabletypesize{\scriptsize}
\tablecaption{USP Planet Radii and Masses\label{tbl-1}}
\tablewidth{0pt}
\tablehead{
\colhead{KIC} & \colhead{KOI} & \colhead{Star Radius ($R_{\odot}$)} & \colhead{Planet Radius ($R_{\oplus}$)} & \colhead{Planet Mass ($M_{\oplus}$)} & \colhead{Distorted Planet Mass ($M_{\oplus}$)}}
\startdata
2718885 &  & $1.454^{+0.053}_{-0.05}$ & $1.551^{+0.131}_{-0.124}$ & $5.169^{+1.912}_{-1.417}$ & $8.116^{+3.468}_{-2.501}$ \\
3112129 & 4144 & $1.012^{+0.028}_{-0.027}$ & $1.142^{+0.045}_{-0.041}$ & $1.67^{+0.244}_{-0.212}$ \\
4665571 & 2393 & $0.741^{+0.016}_{-0.017}$ & $1.299^{+0.041}_{-0.044}$ & $2.661^{+0.315}_{-0.321}$ \\
8435766 &  & $0.757^{+0.015}_{-0.015}$ & $1.25^{+0.026}_{-0.025}$ & $2.312^{+0.181}_{-0.155}$ & $2.624^{+0.201}_{-0.179}$ \\
9642018 & 4430 & $0.827^{+0.021}_{-0.02}$ & $1.349^{+0.073}_{-0.082}$ & $3.052^{+0.653}_{-0.636}$ & $3.884^{+0.803}_{-0.768}$ \\
11187332 &  & $0.983^{+0.03}_{-0.03}$ & $1.203^{+0.072}_{-0.071}$ & $2.012^{+0.468}_{-0.4}$ & $2.393^{+0.544}_{-0.459}$ \\
11550689 &  & $0.647^{+0.013}_{-0.012}$ & $1.435^{+0.031}_{-0.029}$ & $3.839^{+0.338}_{-0.272}$ & $4.513^{+0.383}_{-0.318}$ \\
1717722 & 3145 & $0.669^{+0.015}_{-0.015}$ & $1.193^{+0.053}_{-0.051}$ & $1.952^{+0.332}_{-0.281}$ \\
3444588 & 1202 & $0.667^{+0.015}_{-0.015}$ & $1.453^{+0.05}_{-0.051}$ & $4.044^{+0.529}_{-0.508}$ \\
4055304 & 2119 & $0.812^{+0.017}_{-0.016}$ & $1.383^{+0.032}_{-0.031}$ & $3.362^{+0.289}_{-0.281}$ \\
4144576 & 2202 & $0.856^{+0.017}_{-0.019}$ & $1.188^{+0.032}_{-0.032}$ & $1.922^{+0.201}_{-0.177}$ \\
5040077 & 3065 & $0.956^{+0.025}_{-0.026}$ & $1.203^{+0.076}_{-0.076}$ & $2.014^{+0.498}_{-0.427}$ \\
5095635 & 2607 & $1.003^{+0.028}_{-0.027}$ & $1.69^{+0.073}_{-0.073}$ & $7.213^{+1.329}_{-1.151}$ \\
5175986 & 2708 & $0.654^{+0.016}_{-0.015}$ & $1.612^{+0.047}_{-0.047}$ & $5.99^{+0.735}_{-0.647}$ \\
5513012 & 2668 & $0.894^{+0.021}_{-0.021}$ & $1.53^{+0.042}_{-0.037}$ & $4.895^{+0.531}_{-0.432}$ \\
5942808 & 2250 & $0.788^{+0.018}_{-0.018}$ & $1.753^{+0.054}_{-0.048}$ & $8.364^{+1.123}_{-0.867}$ \\
5972334 & 191 & $0.951^{+0.028}_{-0.028}$ & $1.394^{+0.046}_{-0.052}$ & $3.459^{+0.436}_{-0.461}$ \\
6129524 & 2886 & $0.847^{+0.033}_{-0.03}$ & $1.517^{+0.071}_{-0.073}$ & $4.733^{+0.905}_{-0.79}$ \\
6265792 & 2753 & $1.322^{+0.041}_{-0.041}$ & $1.158^{+0.055}_{-0.052}$ & $1.755^{+0.32}_{-0.272}$ \\
6294819 & 2852 & $1.027^{+0.065}_{-0.058}$ & $1.56^{+0.116}_{-0.099}$ & $5.276^{+1.692}_{-1.153}$ \\
6310636 & 1688 & $1.61^{+0.067}_{-0.066}$ & $1.701^{+0.086}_{-0.086}$ & $7.407^{+1.611}_{-1.37}$ \\
6362874 & 1128 & $0.86^{+0.019}_{-0.019}$ & $1.303^{+0.03}_{-0.028}$ & $2.695^{+0.226}_{-0.217}$ \\
6867588 & 2571 & $1.236^{+0.04}_{-0.039}$ & $1.541^{+0.064}_{-0.062}$ & $5.035^{+0.845}_{-0.715}$ \\
6934291 & 1367 & $0.747^{+0.015}_{-0.015}$ & $1.498^{+0.034}_{-0.032}$ & $4.52^{+0.409}_{-0.342}$ \\
6964929 & 2756 & $1.053^{+0.034}_{-0.031}$ & $1.141^{+0.044}_{-0.042}$ & $1.665^{+0.237}_{-0.213}$ \\
6974658 & 2925 & $0.886^{+0.019}_{-0.019}$ & $0.919^{+0.087}_{-0.092}$ & $0.778^{+0.292}_{-0.237}$ \\
7102227 & 1360 & $0.736^{+0.017}_{-0.016}$ & $0.918^{+0.053}_{-0.053}$ & $0.773^{+0.169}_{-0.14}$ \\
7605093 & 2817 & $0.773^{+0.021}_{-0.021}$ & $1.55^{+0.071}_{-0.065}$ & $5.15^{+0.969}_{-0.773}$ \\
7749002 & 4325 & $0.824^{+0.024}_{-0.023}$ & $1.074^{+0.053}_{-0.055}$ & $1.345^{+0.245}_{-0.228}$ \\
8278371 & 1150 & $1.837^{+0.514}_{-0.303}$ & $1.684^{+0.487}_{-0.276}$ & $7.11^{+13.742}_{-3.53}$ \\
8558011 & 577 & $0.846^{+0.018}_{-0.018}$ & $0.973^{+0.043}_{-0.04}$ & $0.95^{+0.155}_{-0.13}$ \\
8947520 & 2517 & $0.978^{+0.031}_{-0.031}$ & $1.156^{+0.056}_{-0.05}$ & $1.742^{+0.322}_{-0.261}$ \\
9092504 & 2716 & $0.997^{+0.046}_{-0.041}$ & $1.92^{+0.094}_{-0.091}$ & $12.156^{+2.74}_{-2.202}$ \\
9149789 & 2874 & $0.831^{+0.02}_{-0.019}$ & $1.158^{+0.034}_{-0.039}$ & $1.755^{+0.19}_{-0.206}$ & $2.005^{+0.215}_{-0.224}$ \\
9221517 & 2281 & $0.81^{+0.016}_{-0.016}$ & $0.907^{+0.024}_{-0.024}$ & $0.742^{+0.074}_{-0.064}$ \\
9456281 & 4207 & $0.639^{+0.012}_{-0.013}$ & $1.027^{+0.074}_{-0.084}$ & $1.146^{+0.314}_{-0.294}$ \\
9472074 & 2735 & $0.807^{+0.023}_{-0.022}$ & $1.487^{+0.06}_{-0.057}$ & $4.399^{+0.714}_{-0.617}$ \\
9473078 & 2079 & $1.003^{+0.025}_{-0.025}$ & $0.742^{+0.025}_{-0.025}$ & $0.372^{+0.044}_{-0.043}$ \\
9580167 & 2548 & $0.762^{+0.017}_{-0.017}$ & $1.368^{+0.091}_{-0.092}$ & $3.225^{+0.875}_{-0.734}$ \\
10024051 & 2409 & $0.697^{+0.013}_{-0.014}$ & $1.512^{+0.034}_{-0.031}$ & $4.667^{+0.426}_{-0.334}$ \\
10028535 & 2493 & $0.878^{+0.023}_{-0.022}$ & $1.719^{+0.053}_{-0.054}$ & $7.736^{+0.971}_{-0.933}$ \\
10319385 & 1169 & $1.072^{+0.028}_{-0.027}$ & $1.604^{+0.039}_{-0.044}$ & $5.856^{+0.616}_{-0.583}$ \\
10468885 & 2589 & $0.857^{+0.024}_{-0.023}$ & $1.392^{+0.068}_{-0.07}$ & $3.443^{+0.679}_{-0.6}$ \\
10647452 & 4366 & $0.828^{+0.023}_{-0.023}$ & $1.267^{+0.065}_{-0.074}$ & $2.419^{+0.493}_{-0.468}$ \\
10975146 & 1300 & $0.745^{+0.025}_{-0.023}$ & $1.674^{+0.057}_{-0.054}$ & $6.942^{+1.014}_{-0.833}$ \\
11401182 & 1428 & $0.708^{+0.014}_{-0.013}$ & $1.716^{+0.033}_{-0.032}$ & $7.679^{+0.594}_{-0.58}$ \\
11547505 & 1655 & $0.933^{+0.022}_{-0.021}$ & $1.463^{+0.04}_{-0.039}$ & $4.143^{+0.423}_{-0.424}$ \\
11600889 & 1442 & $1.067^{+0.026}_{-0.025}$ & $1.288^{+0.036}_{-0.033}$ & $2.577^{+0.272}_{-0.236}$ \\
11752632 & 2492 & $1.089^{+0.03}_{-0.028}$ & $1.025^{+0.043}_{-0.037}$ & $1.139^{+0.18}_{-0.134}$ \\
11870545 & 1510 & $0.678^{+0.017}_{-0.016}$ & $1.69^{+0.053}_{-0.05}$ & $7.222^{+0.963}_{-0.789}$ \\
11904151 & 72 & $1.07^{+0.027}_{-0.026}$ & $1.553^{+0.042}_{-0.037}$ & $5.19^{+0.543}_{-0.475}$ \\
12265786 & 4595 & $0.793^{+0.019}_{-0.019}$ & $1.416^{+0.115}_{-0.115}$ & $3.651^{+1.26}_{-0.974}$ \\
12405333 & 3009 & $0.804^{+0.018}_{-0.017}$ & $1.086^{+0.054}_{-0.052}$ & $1.396^{+0.267}_{-0.223}$ \\
5773121 & 4002 & $0.846^{+0.02}_{-0.02}$ & $1.431^{+0.042}_{-0.048}$ & $3.798^{+0.457}_{-0.441}$ \\
5980208 & 2742 & $0.619^{+0.011}_{-0.012}$ & $1.012^{+0.035}_{-0.034}$ & $1.093^{+0.135}_{-0.125}$ \\
8416523 & 4441 & $0.767^{+0.015}_{-0.015}$ & $1.399^{+0.033}_{-0.031}$ & $3.499^{+0.307}_{-0.276}$ & $3.965^{+0.356}_{-0.309}$ \\
9353742 & 3867 & $0.971^{+0.027}_{-0.027}$ & $1.746^{+0.053}_{-0.054}$ & $8.234^{+1.085}_{-0.982}$ \\
9475552 & 2694 & $0.751^{+0.014}_{-0.016}$ & $1.504^{+0.031}_{-0.035}$ & $4.586^{+0.378}_{-0.371}$ 
\enddata

\tablecomments{The masses reported the "Planet Mass" in Table \ref{tbl-1} are calculated under an assumption of spherical shape and an Earth-like composition (CMF = 0.33). Distorted masses are reported only for planets with orbital periods less than 10 hours.}

\label{masstable}
\end{deluxetable*}
\end{document}